# “SSL?! What on earth is that?”: Towards Designing Age-Inclusive Secure Smartphone Browsing

**Pavithren V S Pakianathan**
Ludwig Boltzmann Institute for Digital Health and Prevention, Salzburg, Austria
[Corresponding Author]
✉ pavithren.pakianathan@dhp.lbg.ac.at

**L. Siddharth**
Engineering Product Development, Singapore University of Technology and Design, Singapore

**Sujithra Raviselvam**
Industrial Design, Eindhoven University of Technology, Eindhoven, Netherlands

**Kristin L. Wood**
College of Engineering, Design and Computing, University of Colorado Denver, CO, USA

**Hyowon Lee**
School of Computing, Dublin City University, Dublin, Ireland

**Pin Sym Foong**
Saw Swee Hock School of Public Health, National University of Singapore, Singapore

**Jianying Zhou**
Information Systems Technology and Design, Singapore University of Technology and Design, Singapore

**Simon Tangi Perrault**
Information Systems Technology and Design, Singapore University of Technology and Design, Singapore

## Abstract

Owing to the increase in ‘certified’ phishing websites, there is a steady increase in the number of phishing cases and general susceptibility to phishing. Trust mechanisms (e.g., HTTPS Lock Indicators, SSL Certificates) that help differentiate genuine and phishing websites should therefore be evaluated for their effectiveness in preventing vulnerable users from accessing phishing websites. In this article, we present a study involving 18 adults (male-6; female-12) and 12 older adults (male-4; female-8) to understand the usability of current trust mechanisms and preferred modalities in a conceptualized mechanism. In the first part of the study, using Chrome browser on Android, we asked the participants to browse a banking website and a government website for digital particulars. We asked them to identify which one of the two was a phishing website, rate the usability of both websites and provide qualitative feedback on the trust mechanisms. In the second part, we conceptualized an alternative trust mechanism, which allows seeking social, community and AI-based support to make website trust-related decisions. Herein, we asked the participants as to which modality (social, community or AI) they prefer to seek support from and why it is preferred. Using the current trust mechanisms, none of the participants were able to identify the phishing website. As the participants rated the current mechanisms poorly in terms of usability, they expressed various difficulties that largely did not differ between adults and older adults. In the conceptualized mechanism, we observed a notable difference in the preferred modalities, in that, older adults primarily preferred social support. In addition to these overall findings, specific observations suggest that future trust mechanisms should not only consider age-specific needs but also incorporate substantial improvement in terms of usability.

## 1. Introduction

Being the weakest links in cybersecurity [1], internet users are often prone to phishing, which is a form of social engineering technique aimed at credential and personal identity theft (e.g., bank account details). As reported globally, internet users are susceptible to phishing through websites, emails, messaging services, phone calls, social media platforms, and other applications. Despite studying, testing, and implementing various phishing interventions, several users, especially older adults, are still susceptible to phishing. We argue that these interventions are yet to address age-related difficulties and preferences. In the current version of an Android web browser, to check the authenticity of a website, e.g., a banking website shown in Figure 1, a user may examine the URL (Figure 1a) to verify the domain name – "https://internet-banking.dbs.com.sg". The lock symbol (🔒) indicates a secure connection (Figure 1B), whose type and details could be verified as indicated in Figure 1c and Figure 1d. An insecure connection is indicated as information required (ⓘ) or dangerous (▲). Due to the increase in the number of certified phishing websites, these indicators need not be sufficient to ensure the website's authenticity [2]. Instead, these indicators merely ensure that communication between the user and the website remains private and unaltered.

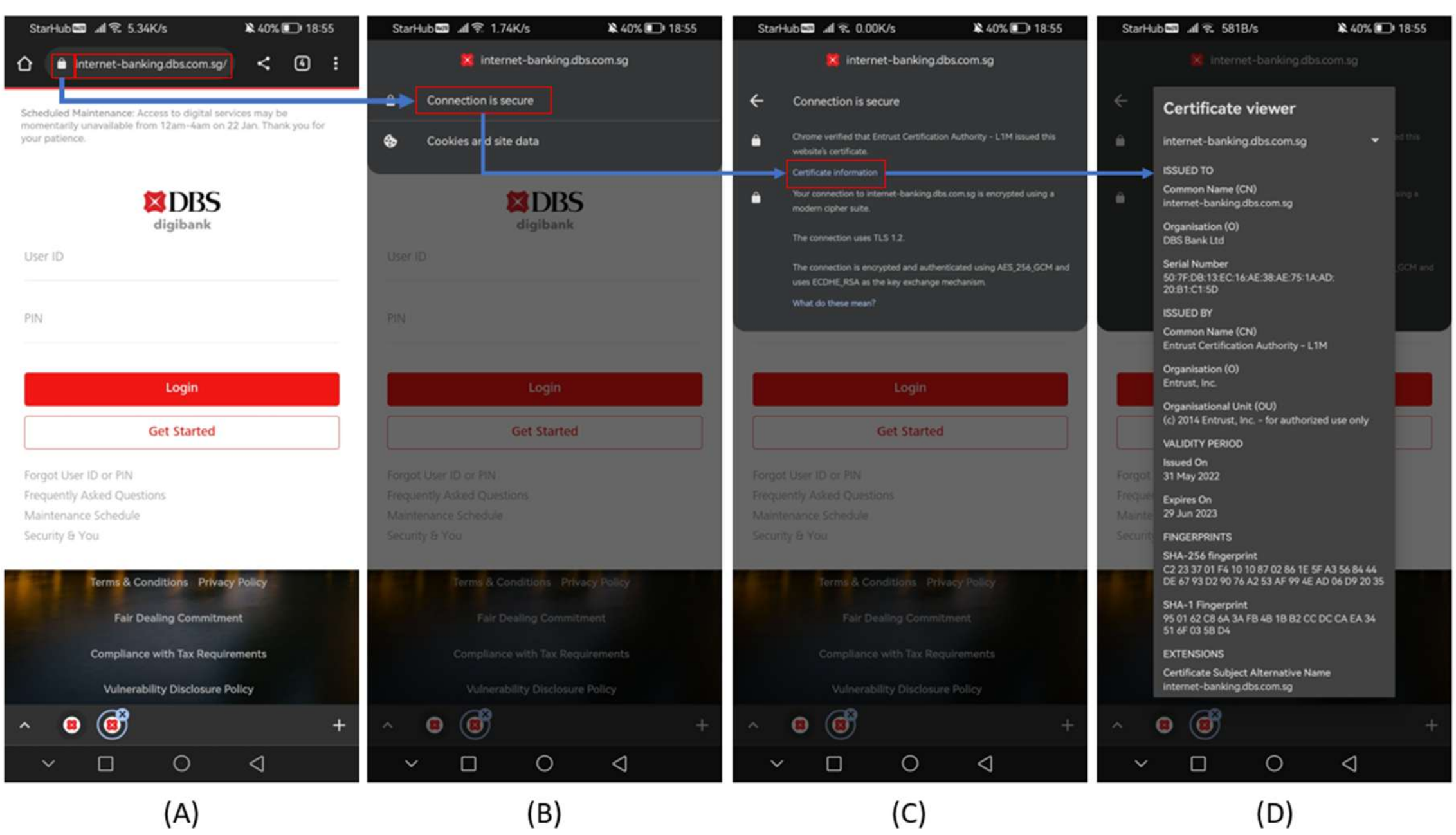

Figure 1: Implementation of trust mechanisms in an Android Chrome Browser

Various technology-mediated social interventions (e.g., social-proof and community-based support) have been proposed to improve resilience to phishing. Subject to availability [3], older adults vastly prefer social interventions such as family, friends [4], and occasionally community [5–7]. Literature also indicates that social interventions significantly influence learning and decision-making in cybersecurity [5,7–9]. In addition, the literature also reports the growing impact of AI-based interventions which have relatively high accuracy - about 90% or higher [10,11]. Unlike social interventions, which depend on the knowledge and capabilities of a limited user group, these AI-based interventions primarily rely on machine learning-based models which are trained on large datasets. Carrying individual advantages and limitations, these different interventions should therefore complement each other. Phishing susceptibility has often been linked to user characteristics such as age, gender, special needs (e.g., visual impairments), technology awareness, cognitive ability, and others [12–15]. Despite the increase in attention received for inclusive design solutions [16,17], the needs experienced by older adults based on certain physical and cognitive challenges are rarely considered while developing phishing interventions [18]. While several studies examine the effectiveness of implementations of security indicators and alerts across different apps and devices [19,20], it is less understood as to whether these implementations are valuable and usable in adherence to user characteristics such as age. Age is an umbrella characteristic that is often linked to other factors, such as technology awareness and physical and cognitive ability. Hence it is essential to examine the usability and usefulness of the security indicators based on the user's age.

## 2. Background

### 2.1. Awareness-raising interventions

According to Franz et al. [18], awareness-raising interventions communicate risks to the users while, for example, opening an email. These interventions could be active or passive, depending on whether a user must perform an action in response. SSL certificates and lock icons are considered passive interventions. Drury and Meyer [2] found that certificates (as in Figure 1D) rarely help distinguish between legitimate and phishing sites. In contrast, Gopavaram et al. [13] found participants to be reliant on lock icons (as in Figure 1A) for identification of legitimate websites, provided they are young and have adequate computer experience and a decrease in age. Among active interventions, Volkamer et al. [21] proposed the TORPEDO plugin for email applications, wherein, the weblinks pop up the entire URL with the domain highlighted along with necessary tooltips that delay the access to the URL. They found it to be more effective compared to merely examining the URL in the status bar of a web browser.

While studies suggest that current security indicators based on domain and certificate are less effective forms of interventions [22] that are often misunderstood, ignored, or unnoticed [19,20] interactive warnings are effective [18] as well as preferred by the users [23]. Furthermore, it has been shown that, compared to desktop usage, browsing via smartphone leads to three times as likely to access phishing sites, mainly due to device-specific limitations such as reduced screen real estate and lack of security indicators [24,25]. With mixed findings on the relationship between age and phishing [26,27], it is necessary to examine the age-related difficulties and associated challenges that stem from the lack of technical expertise, and physical and cognitive limitations.

### 2.2. AI-based Interventions.

AI-based interventions adopt state-of-the-art classifiers to detect phishing websites via URL, domain, certificate, and context surrounding the websites [28]. While such interventions are quite effective in terms of preventing users from being exposed to phishing content via email clients and messaging applications, the users are still susceptible to phishing while browsing through various websites due to occasional false positives. For example, in the Android Chrome browser, a full-screen red warning page is presented when a user visits an unsafe website. This is triggered automatically based on Google's automated systems. While the technical performances of AI-based interventions continuously improve, the trust towards such interventions is yet to be examined in the literature [28–30].

### 2.3. Social Interventions.

Users, especially older adults, often learn about cybersecurity and/or acquire help thereof from their social circle [3,7,31–34]. A lack of social support, especially for older adults, could result in higher susceptibility to phishing. Based on this premise, Chouhan et al. [5] developed the "Community Oversight for Privacy and Security" model, which guides users seeking social help for privacy and security decisions. Studies using their model indicated that users are likely to receive help primarily from close friends and family members. Wan et al. [7] developed the "App-ModD" approach to obtaining security help from the community. The app-based implementation of their approach improved the accuracy of security decisions made by older adults. Burda et al. [35] observed that In addition to community-based assistance, individuals who can detect cyber anomalies help improve an organization's resilience to phishing.

Apart from seeking security-related help, users from both age groups adopt secure browsing habits and acquire security knowledge among social network members [9]. Similarly, stories or facts shared by peers serve as an effective form of phishing training [31,32]. Poole et al. [3] found that individuals rely on social help based on availability rather than technical capability. Since social support might not be readily available, it might be important to assess how other modalities of support could act as fallback methods in case individuals require support with making trust decisions.

### 2.4. Ageing vs. Phishing

Frik et al. [36] posited that technical capabilities and experiences are influential factors in phishing. Despite being more suspicious of phishing, older adults have increased susceptibility to getting phished or defrauded, primarily due to the decline of general cognitive ability, sensitivity to deception [37], and lack of information about fraud prevention [38]. Grilli et al. [26] found that older adults are more likely to be victimised by online fraud due to reduced sensitivity to the credibility of emails. Gavett et al. [12] found that older adults' suspiciousness of phishing was explained by their educational background and prior phishing knowledge. Furthermore, older adults do not often rely on the internet for seeking phishing-related information [39]. However, Gopavaram et al. [13] found a negative correlation between age and phishing detection, indicating that younger adults could also be vulnerable to phishing.

As influencing factors of phishing, prior studies examine user characteristics such as age, technology experience, gender, cognitive ability, and exposure to phishing training [13,14,40]. Prior studies also distinguish active vs. passive warnings [41], suggesting that the former are more effective intervention mechanisms. However, these studies seldom take into account the usability of the interventions in reference to the aforementioned user characteristics. In particular, these studies do not focus on age-related difficulties in the context of browsing the internet using a smartphone. As older adults are increasingly adopting smartphones and are considered vulnerable to phishing, it is important to evaluate the effectiveness of such interventions. In these lines, we ask the following research questions.

*RQ 1: What are the age-related difficulties while using trust mechanisms in current smartphone browsers?*
*RQ 2: What are age-related preferences towards social-, community-, and AI-based supports to assess the trustworthiness of websites?*

To address the questions stated above, we conduct a study as shown in Figure 2. Upon capturing the demographic information (as listed in Table 1) from participants, for RQ 1, we assess the usability of the current trust mechanisms in a Chrome browser based on Android. For RQ 2, we conceptualise a trust mechanism that acquires support from various modalities and evaluate it during the second part of the study. We then acquire some opinions from the participants regarding their browsing habits.

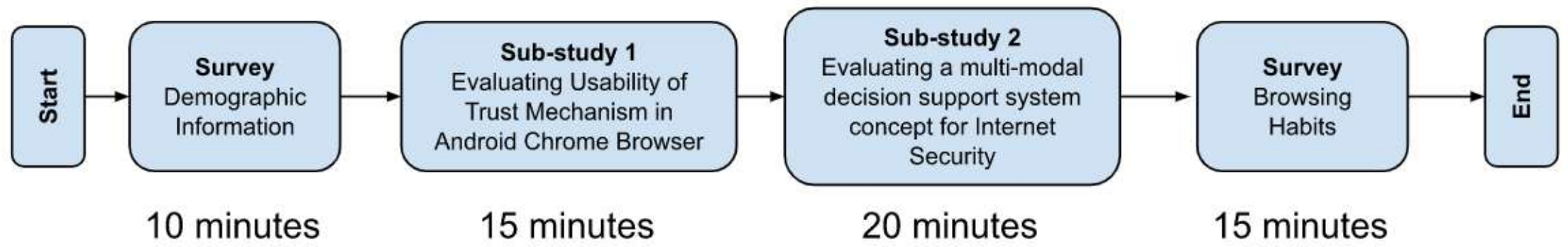

Figure 2: Study Flow

## 3. Method

### 3.1. Pre-Studies

Prior to this study, we conducted semi-structured interviews with older adults (n=10) and caregivers (n=10) to identify how older adults navigated and received support privacy and security. Subsequently, we conducted an online survey (n=50) to identify the following key issues related to privacy and security: 1) unaware of permissions when installing applications on smartphones, 2) difficulties in password management, and 3) poor understanding of URLs. Based on these issues, we conducted six online workshops (n=18) with experts from HCI, security and design backgrounds to ideate necessary features that are required to be implemented in trust mechanisms.

Upon ideating features through "how might we" statements, we sorted these features using affinity diagrams and filtered the following: 1) support multiple modalities for getting help and trust-related feedback (family members, online community/crowdsourcing, automated tools/AI), 2) warning notifications for malicious websites, 3) allowing rating of members in the community, 4) safe mode, 5) trust score for each website, and 6) forced attention. Based on these six features, we focussed on the most important feature – support for multiple modalities and conceptualised an application that we evaluate in the second part of our study.

### 3.2. Study Overview

As depicted in Figure 3, the study involved collecting demographic information, understanding age-related difficulties associated with current smartphone browsers, evaluating the conceptualized trust mechanism (Figure 4), and gathering opinions on browsing habits. Through Telegram advertisements (reward $10 shopping vouchers), we recruited participants representing older adults (n=12) and adults (n=18) in a Southeast Asian city. Except for one older adult, all participants have been using a smartphone for more than 3 years. Adult participants were 18 to 64 years old (12F, 6M), and older adult participants were above 65 (8F, 4M). Eight participants used an iPhone, and the remaining 22 used an Android smartphone.

Table 1: Participant Demographic Information

| # | Gender | Age | OS | Education level | Occupation | Phishing |
|---|---|---|---|---|---|---|
| 1 | Male | 55-64 | Android | O level | Money Changer | 4 |
| 2 | Female | 25-34 | iPhone | Polytechnic Diploma | Student | 1 |
| 3 | Male | 25-34 | Android | Undergraduate Degree | Software Developer | 4 |
| 4 | Male | 25-34 | Android | Polytechnic Diploma | Art Director | 1 |
| 5 | Female | 25-34 | Android | Undergraduate Degree | Educator | 2 |
| 6 | Female | 18-24 | iPhone | Undergraduate Degree | Public Servant | 2 |
| 7 | Female | 18-24 | Android | A Level | Student | 3 |
| 8 | Female | 35-44 | Android | Undergraduate Degree | Accountant | 3 |
| 9 | Male | 55-64 | Android | Polytechnic Diploma | Course Consultant | 3 |
| 10 | Female | 65-74 | iPhone | Postgraduate Degree/Doctorate | Retired Teacher | 3 |
| 11 | Male | 65-74 | iPhone | Undergraduate Degree | Technician | 1 |
| 12 | Female | 55-64 | Android | Undergraduate Degree | Caregiver | 4 |
| 13 | Female | 55-64 | Android | Other | Gig Job Worker | 3 |
| 14 | Female | 65-74 | Android | Undergraduate Degree | Self-Employed | 4 |
| 15 | Female | 65-74 | Android | Primary School | Self-Employed | 4 |
| 16 | Female | 55-64 | Android | A Level | Personal Assistant | 2 |
| 17 | Male | 35-44 | Android | Postgraduate Degree/Doctorate | Technology Consultant | 4 |
| 18 | Female | 45-54 | Android | O level | Stocks Checker | 2 |
| 19 | Male | 65-74 | Android | Polytechnic Diploma | Freelancer | 3 |
| 20 | Male | 65-74 | iPhone | O level | Exam Invigilator | 1 |
| 21 | Female | 65-74 | Android | O level | N.A. | 2 |
| 22 | Female | 75 < | Android | Postgraduate Degree/Doctorate | Website Designer | 4 |
| 23 | Male | 75 < | iPhone | O level | N.A. | 2 |
| 24 | Female | 55-64 | iPhone | A Level | Unemployed | 2 |
| 25 | Female | 75 < | iPhone | Polytechnic Diploma | Retired Teacher | 1 |
| 26 | Female | 65-74 | Android | Undergraduate Degree | Retired Civil Servant | 1 |
| 27 | Female | 65-74 | Android | O level | Gig Job Worker | 1 |
| 28 | Male | 25-34 | Android | Undergraduate Degree | Student And Consultant | 4 |
| 29 | Female | 25-34 | Android | Undergraduate Degree | Teacher | 3 |
| 30 | Female | 25-34 | Android | Undergraduate Degree | Quality Assurance Engineer | 2 |

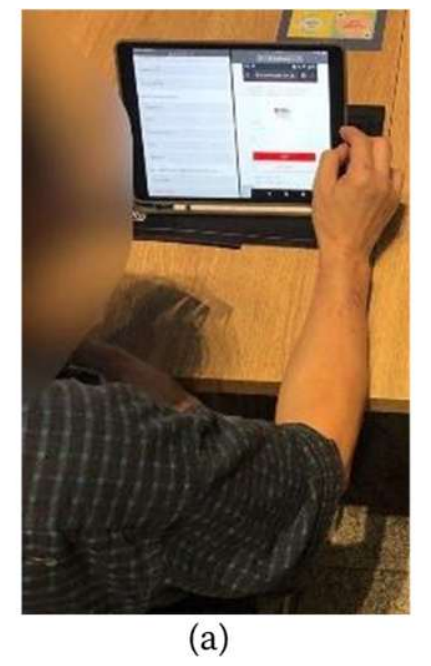

(a)
A participant taking part in the study with Qualtrics form and application prototype

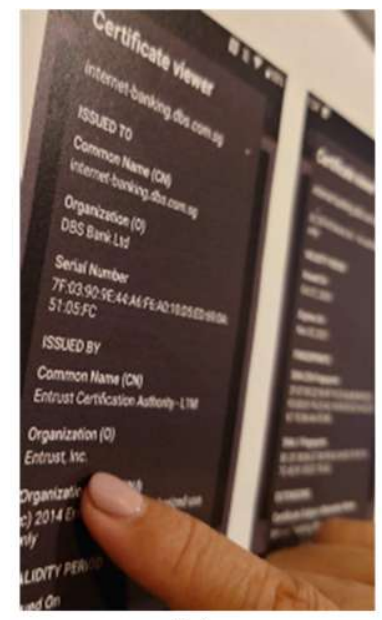

(b)
"I don't know Entrust" - P28, 25-34 y.o (Genuine DBS website)

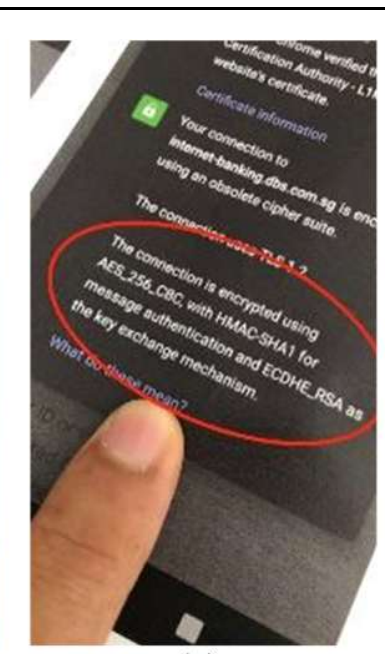

(c)
"Normal person won't know [content within oval]" - P26, 65-74 y.o. (Genuine DBS website)

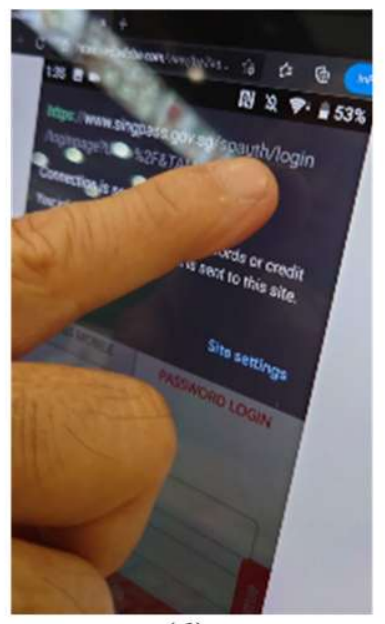

(d)
"/spauth looks suspicious" - P19, 65-74 y.o. (Genuine DBS website)

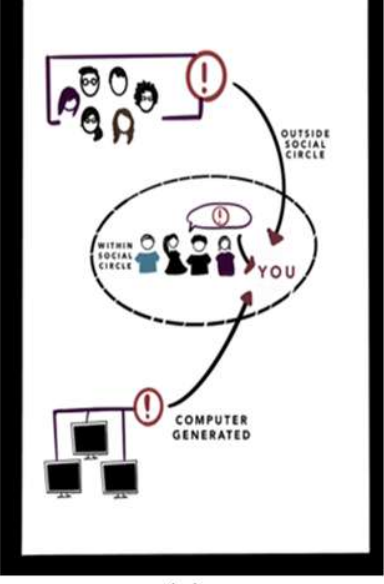

(e)
A printout was given to participants to explain how the design concept works

Figure 3: Pictures captured during the study and supplementary material.

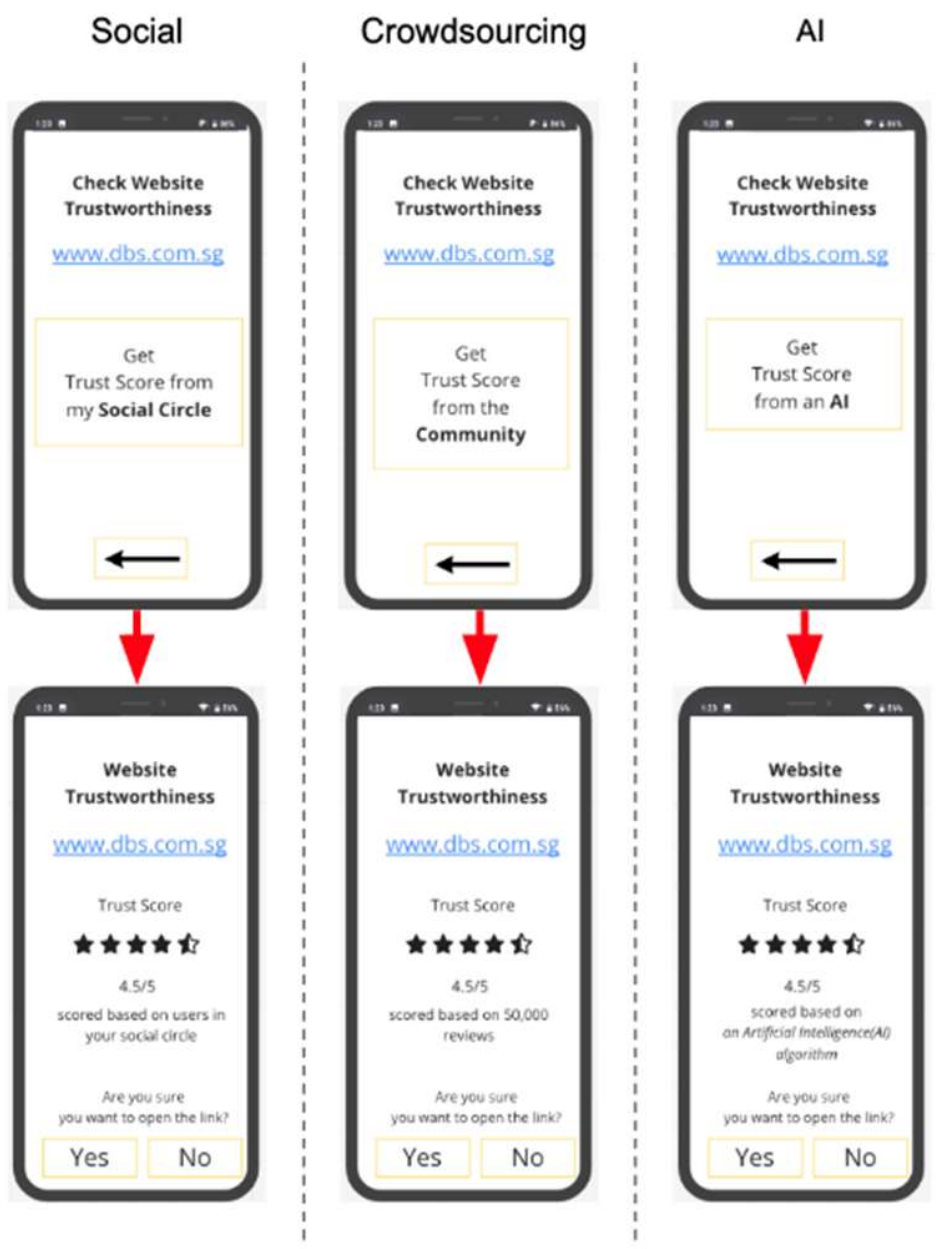

Figure 4: Conceptualized Trust Mechanism.

We assessed their web-use skills [42] based on the understanding of the following terms: PDF, Spyware, Wiki, Cache and Phishing. Both age groups had medium to little understanding of the term phishing: (older adults: mean = 2.25, SD=1.29; adults: mean = 2.72, SD=1.02). We have provided the details of individual participants in Table 1. We have detailed the intermediate steps, analysis methods, and outcomes of our study in Table 2. In the first of the study, to identify the age-related difficulties with current smartphone browsers, we developed a clickable wireframe using AdobeXD. We have provided accessible links to these prototypes in APPENDIX I and indicated the URL part of the prototype in Figure 5. We developed the prototypes for a local banking website and a government website for digital particulars that involve highly sensitive information. The prototypes include an Android Chrome browser (due to high usage) and the associated trust mechanisms such as the website certificate details and URL (Figure 5).

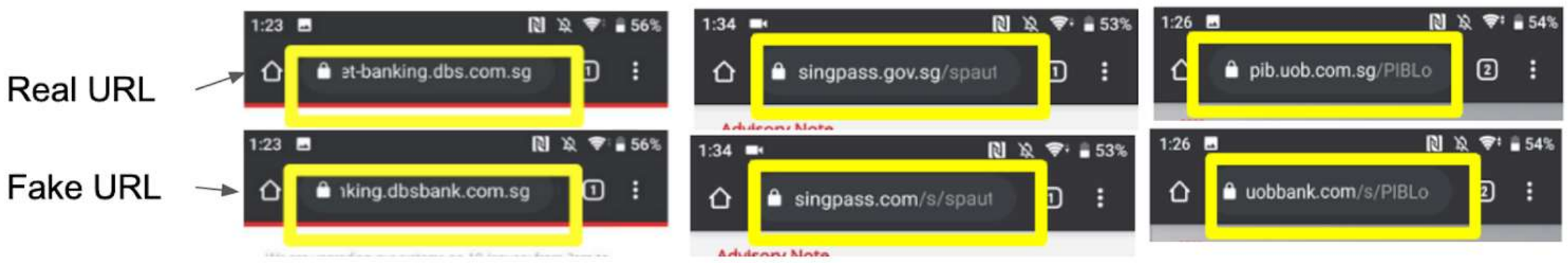

Figure 5: Screenshots of URL bar with subtle differences in URLs

As mentioned in Table 2, we asked the participants to rate their familiarity with the websites and interact with two prototypes. We asked them to imagine themselves logging into these websites, given that one of the two could be fake. We particularly asked them to "think aloud" while using the trust mechanisms. When participants did not click on the lock symbol or view the full URL or the certificate details, we prompted them to continue exploring till all the pages are exhausted. Upon exploration, we asked them to identify whether the website is fake or genuine. We then asked them to rate the usability of trust mechanisms using the System Usability Scale (SUS). Throughout the think-aloud activity, we kept note of their activities and opinions. As shown in Figure 4, we conceptualized an alternative trust mechanism and presented it to the participants on a printed A4 sheet along with a supporting diagram shown in Figure 3e. As mentioned in Table 2, we asked the participants to state the preferred modality based on the following inputs: 1) *"I would submit a URL for verification to…"* and 2) *"I would trust the feedback of…"* We then asked them to state their preferences using a 5-point Likert scale and provide a short justification for their response.

Table 2: Study Overview

| Participant Activity (Duration) | Purpose | Participant Output | Analysis Methods | Analysis Outcomes |
|---|---|---|---|---|
| Survey<br><br>(10 min) | To gather demographics and technology experience | Objective Responses | - | Demographics and web-use skills of participants. |
| **Sub-Study I**<br>1) Rate familiarity with websites<br>2) Interact with the websites<br>3) Identify the fake one<br>4) Rate usability<br>5) Provide qualitative feedback on usability<br><br>(15 min) | To assess age-related difficulties | Observation notes, Usability scores, Descriptive Response (Qualitative feedback) | **Quantitative Analysis**<br>Analysis of familiarity ratings, Analysis of usability scores, according to the SUS scale<br><br>**Qualitative Analysis**<br>Extraction of key topics, Development of codebook, Interrater reliability, Refinement of the codebook, Analysis of coding frequency | The number of adults and older adults who correctly identify fake websites.<br><br>Usability grades (w.r.t., SUS scale) for current trust mechanisms according to adults and older adults.<br><br>Topics (with frequencies) identified from the qualitative feedback on usability from adults and older adults. |
| **Sub-Study II**<br>1) Understand the conceptualized trust mechanism<br>2) Select the modality using the Likert scale<br>3) Justify the preferred modalities<br><br>(20 min) | To assess age-related preferences of modalities | Objective response (modality preference), Descriptive responses (justification) | **Quantitative Analysis**:<br>Analysis of modality preferences<br><br>**Qualitative Analysis**:<br>Extraction of key topics, Development of codebook, Interrater reliability, Refinement of the codebook, Analysis of coding frequency | Preferred modalities of adults and older adults.<br><br>Topics (with frequencies) identified from the justifications given by adults and older adults regarding the preferred modalities. |

## 4. Results

As mentioned in Table 2, we collected both quantitative and quantitative responses (in local colloquial language and highlighted in Gray) from the participants. To analyse the qualitative data, we invited three researchers to tag the responses using codes (listed in **APPENDIX II**). Based on the responses from the first 10 participants we observed a Fleiss' $\kappa = 0.75$, indicating a satisfactory agreement of codes assigned. After resolving the disagreements, one of the researchers coded for the remaining 20 participants' responses.

### 4.1. Sub-Study 1

Participants from both age groups expressed high familiarity with the websites. Specifically, for the local government website, adults expressed 4.22 on average (SD = 0.73) while older adults expressed 3.92 (SD = 0.9). Similarly, for the banking website, adults gave 4.22 (SD = 0.73) while older adults gave 3.67 (SD = 0.65). Overall, older adults rated lesser familiarity than adults. None of the participants, from both groups, could correctly identify the fake website. As mentioned by Hasegawa et al. [43], security concepts are generally complex for a non-expert for users to grasp. However, even participants with technical experience (a software developer (P3) and an IT consultant (P17)) could not detect fake websites. This also concurs with a low understanding of "Phishing" among participants as shown in Table 1.

When asked to rate the usability according to SUS Scale, the adults rated 45.29 on average while older adults rated 28.96. According to the SUS Scale, scores could be interpreted as follows: > 80.3 (A - Excellent), 68 – 80.3 (B - Good), 68 (Okay), 51 – 68 (D - Poor), < 51 (F - Awful). Based on these interpretations, both groups rated the usability as 'Awful' and older adults rated usability significantly poor compared to adults. These observations suggest that trust mechanisms require overall improvement while also taking into account age-specific needs. Upon examining the qualitative responses and codes assigned to these (summarized in Table 3), we discuss the age-related difficulties according to various aspects as follows.

Table 3: Codes representing the age-related difficulties with current trust mechanisms.

| Adults | Older Adults |
|---|---|
| **Issues with current trust mechanisms** | |
| **lack of awareness (15)**, incomplete information (2), habituation (4), **mismatch in information provided (8)**, misconception (4), lack of preventive measures (1), vague (2), too much information (0), anxiety (1) | **lack of awareness (7)**, incomplete information (0), habituation (0), **mismatch in information provided (22)**, misconception (0), lack of preventive measure (0), vague (2), too much information (2), anxiety (0) |
| **Methods for Accessing and Trusting** | |
| avoidance (7), data sensitivity (5), source validity (4), takes preventive measures (1), vague (1) personal device (2), **checks security indicator (15), checks aesthetics of website (7), trust towards entity (8)**, additional security (1), check with social circle (2), lack of preventive measures (5), check online (2) | avoidance (1), data sensitivity (1) source validity (1), takes preventive measures (0), vague (1), personal device (4), **checks security indicator (6), checks aesthetics of website (10), trust towards entity (6)**, additional security (2), check with social circle (5), lack of preventive measures (0), check online (2) |

***Issues with current trust mechanisms***. As shown in Table 3, participants reported a lack of awareness and a mismatch in the information provided, predominantly among older adults. For example, participants were not even aware of the SSL certificate (Figure 3b) as illustrated below.

*"SSL?! What on earth is that?" - Older Adult, P25*

Specifically, in terms of SSL certificates, older adults expressed that the information is overwhelmingly technical. When clicking on "what do these mean" (Figure 4c), participants expressed the following concerns.

*"The certificate is so technical - I need clarification", Older Adult, P27*
*"The security info sounds non-sensical…rubbish!", Older Adult, P11*
*"Not cumbersome [to understand], but the information presented is not useful", Adult, P28*

***Methods for accessing and trusting***. As shown in Table 3, to assess the trustworthiness of a website, participants expressed that they primarily checked the security indicators and aesthetics of the website. For instance, an older adult only checked the "logo or appearance" to assess trustworthiness (P10). Another older adult misjudged the website by spotting the folder name (Figure 3d) in the URL. While adults checked security indicators more frequently, older adults typically examine the aesthetics as expressed below.

*“[I look for] HTTPS and [do] not usually scrutinise the URL)”, Adult, P12*
*“HTTPS means secure”, Adult, P13*
*“[The] lock is not really useful, so [I] the appearance of the website to determine if the website is real or fake”, Adult, P2*
*[I don’t know] what the lock implies, [I] thought it is used to block websites, Older Adult, P10*

As shown in Table 3, another common method to identify legit websites is based on the trust toward the entity, i.e., the implicit trust towards the hosting organisation as illustrated below.

*“[I] trust the government website”, Adult, P1*
*“[A] government website cannot be non-reliable” Older Adult, P27*

The overall findings from Sub-Study I are that none of the participants was able to identify the fake website and rated ‘Awful’ usability for the trust mechanisms, despite mentioning high familiarity with the websites. Based on the common themes that emerged from the qualitative responses, both groups expressed awareness and mismatched information provided (more enhanced for older adults). Both groups utilise common methods such as security indicators, aesthetics and hosting organisations to assess trustworthiness. Although some of these themes are more enhanced for either of the groups, the difficulties expressed by the participants largely do not differ with age, indicating that trust mechanisms require an overall improvement, while also taking into account age-specific needs.

### 4.2. Sub-Study II

In the second part of the study, we collected the modality preferences based on two inputs. To examine whether age played a role in their preferences, we present a 1-way ANOVA in Table 4. In terms of acquiring support from social circles, our results indicate a statistically significant difference between the groups, i.e., older adults prefer social circles considerably higher than adults. Similar to the first part of the study, we organized the codes assigned to the qualitative responses and compared these across groups in Table 5.

Table 4: ANOVA to assess preference towards modalities.

| **Response** | **Mean Ratings** | | **Results** |
|---|---|---|---|
| | **Adults (n=18)** | **Older Adults (n=12)** | |
| Willingness to submit URL to Social Circle for verification | 2.44 | 4.08 | $F(1,27) = 13.76$, $p = .0009$ |
| Willingness to submit URL to outside Social Circle for verification | 3.12 | 3 | n.s. ($p = .82$) |
| Willingness to submit URL to AI for verification | 3.29 | 3.83 | n.s. ($p = .19$) |
| Trust feedback of Social Circle | 2.94 | 4 | $F(1,27) = 6.46$, $p = .02$ |
| Trust feedback of outside Social Circle | 3 | 2.58 | n.s. ($p = .40$) |
| Trust feedback of AI | 3.71 | 3.75 | n.s. ($p = .91$) |

Table 5: Codes representing preferences in modality.

| **Adults** | **Older Adults** |
|---|---|
| **Submit to Social Circle** | |
| **trust reliability of helper (12)**, independent (1), **contextual preference (4)**, **cannot trust reliability of helper (9)**, time taken to get response (3), **availability of helper (9)**, causes trouble for helper (1), vague (2), lack of preventive measures (0), cause trouble for helper (1) | **trust reliability of helper (21)**, independent (2), contextual preference (1), **cannot trust reliability of helper (4)**, time taken to get response (0), **availability of helper (4)**, causes trouble for helper (2), vague (0), lack of preventive measures (1), cause trouble for helper (0) |
| **Submit to Outside Social Circle** | |
| **cannot trust reliability of helpers (18)**, **independent (7),** availability and quick response (1), **potential manipulation (5)**, **wisdom of the crowd (7)**, trusted platform (1), unfamiliar method (0), vague (1), reliability of scoring mechanism (2) | **cannot trust reliability of helpers (10)**, independent (0), availability and quick response (1), **potential manipulation (4)**, **wisdom of the crowd (5)**, trusted platform (4), unfamiliar method (1), vague (1), reliability of scoring mechanism (0) |
| **Submit to AI** | |
| Unfamiliar method (1), **AI dependability (27)**, **potential manipulation (5)**, availability and quick response (1), combination (1), vague (2) | Unfamiliar method (2), **AI dependability (16)**, **potential manipulation (3)**, availability and quick response (0), combination (2), vague (1) |

***Submit to social circle***. In terms of trusting the reliability of social circles, participants expressed opposite views, while older adults largely seem to trust the reliability as illustrated below.

*"If it's within my social [circle], I also trust my friends, people that I know and value their opinions", Older Adults, P22*

*"I would send it to my daughter. She is [an] IT specialist and knows better than me.", Older Adult, P22*

As shown in Table 5, adults have particularly expressed contextual preference, i.e., acquiring help based on the technical capabilities of the helper.

*"I would trust my inner social circle since some of them have backgrounds in cybersecurity and are more tech-savvy", (Adult, P3)*

Common to both groups, participants, especially adults, have expressed concerns about the availability of the helper in order not to cause trouble. The availability also meant that no one exists in the social circle to provide help.

*"[They] could be busy, or it might be too troublesome for them" Adult, P18.*

*"I do not have immediate circle or friends or family to help me", Adult, P13*

***Submit outside the social circle***. In terms of submitting outside social circle, i.e., to the community, participants have largely expressed concerns regarding the reliability and manipulation of helpers.

*"I dunno whether they will sabo[tage] me. [I]'m very scared", Older Adult, P27*

*"It has to be a trustworthy forum with active users", Adult, P8*

*"They are stranger[s] to me. I dunno whether they are the expert or not. I will need to see a review of the user. But [the] review can be fake", Adult, P9*

As the participants find it concerning to acquire help from the community, as shown in Table 5, adults take a step further and rely on their own technical abilities to assess the trustworthiness of websites.

*"Friends in my social circle are around my age, so they are not tech-savvy. They have to ask me for help." Adult, P24*

Meanwhile, some felt that they make the assessment themselves.

*"I would be able to validate the website myself. If I can't, I will not use that website", Adult, P3*

*"If something is too good to be true, I will google it and check online if it's something related to a scam or something," Adult, P17*

While concerns about manipulation are common across groups, an older adult mentioned that community input could complement other modalities.

*"Good to get [crowd workers'] feedback as [an] additional check", (Older Adult, P14)*

***Submit to AI***. Regarding AI, participants generally question the dependability, expressing concerns about manipulation.

*"Not sure AI is reliable or not and depends on context", (Adult, P30)*

*"I got trust issues with the chatbots and AI", (Adult, P30)*

*"I still don't think AI is there, and I still feel that I need verification from humans on whether to trust what the AI put out", (Adult, P28)*

Similar to how participants relied on trustworthy organisations to judge a website, they trust AI if it is developed by trustworthy and reliable organisations.

*"[I will] trust the person [who] created the AI - [if it's] created by gov[ernment], [a] well-known company, or university", (Older Adult, P21)*

Overall, from the modality preference, it appears that older adults significantly prefer support from social circles compared to adults. The specific reasons behind their preferences indicate that older adults largely trust the reliability of helpers while also expressing concerns about their availability. Adults, in particular, have shown a contextual preference towards social circles in that they choose help based on technical capabilities and availability. Participants have generally expressed aversion towards acquiring help from the community, stating concerns over reliability and manipulation. These responses have also uncovered the adults who are independent and rely on their own technical capabilities. While the participants have largely questioned the dependability of AI, they are inclined to acquire help if it is built by trustworthy organizations.

## 5. Discussion

### 5.1. Design Implications

The key issues with current trust mechanisms, as identified from Sub-Study I, are a lack of awareness and a mismatch in the information provided. While terminologies are quite technical, it is questionable whether users access the "what do this mean" links in certificates. Android browser developers could integrate cybersecurity training and awareness within browsers via just-in-time tooltips [21] and localized phishing cases. However, integrating such features without hindering user experience is still a challenge and would need to be further explored. As participants check aesthetics in common, web developers shall not compromise on this aspect. Despite utilizing security indicators, none of the adults was able to identify the fake websites. While prior studies have shown that domain highlighting, lock symbols and certificates are ineffective, developers could rethink how URLs could be represented (Figure 6) such that phishing websites could be identified.

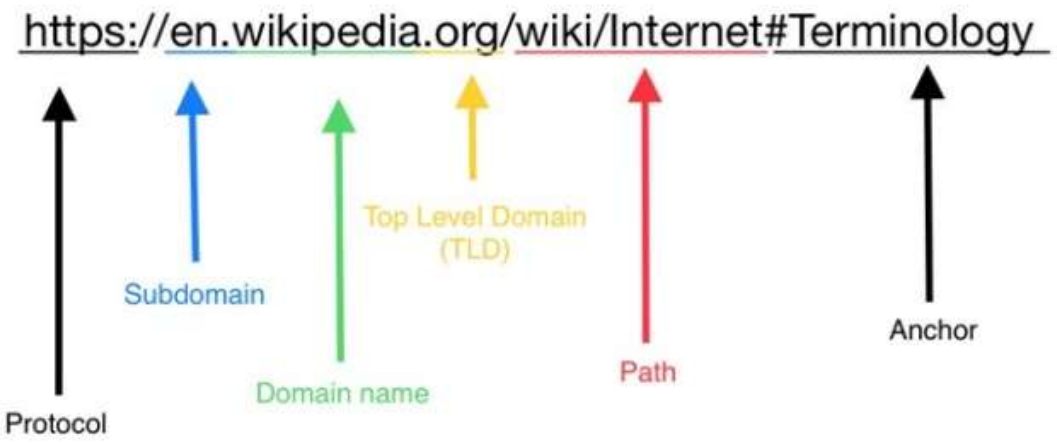

Figure 6: URL Structure

Support from social circles seems to be a much-needed feature in future trust mechanisms, especially for older adults. According to the preferences of adults, it would be appropriate to indicate the technical capabilities and the availability of the helper as well. Community support largely does not seem to be preferred, although it could be complemented by other modalities. Developers could therefore minimize efforts into interconnecting large communities and focus more on integrating the opinions of the inner social circle (e.g., frequently contacted) in the browsers. As few adults tend to be independent, the developers could accommodate different kinds of information such that tech-savvy adults make informed decisions. As far as AI is concerned, more than expressing concerns, participants were questioning the dependability. Specifically, they were ready to seek help from AI if it is developed by a trustworthy organization, which is likely to have implemented the recommendations of trustworthy AI [29]. The developers could therefore integrate state-of-the-art AI models from renowned companies like Microsoft and make sure the company names are communicated to the user.

### 5.2. Limitations

As tech experience is self-reported by the participants it could have occurred that a few participants had relatively lesser or more understanding of security indicators and modalities in the study. Since the study is conducted in the Asian context, it might not be representative of the population that is relatively less or more tech-savvy. The older adult participants were more technologically enthusiastic than expected. If otherwise, the concerns expressed by older adults in Tables 3 and 5 would be more enhanced. Although sub-study II does not include an actual prototype, the responses from participants have given some ideas regarding the modalities and features to be implemented in actual browsers. In the future, we intend to study the age-related differences with an actual prototype, while also addressing the aforementioned limitations and recruiting a more diverse set of participants.

## APPENDIX I

| Entity | Original website | Fake website |
|---|---|---|
| DBS | https://xd.adobe.com/view/bde549f6-4613-4996-8742-5c69de042a49-e8b6/?fullscreen&hints=off | https://xd.adobe.com/view/ac3c69d5-3f2a-48db-8a61-a4a620554669-f96f/?fullscreen&hints=off |
| UOB | https://xd.adobe.com/view/7541f4d5-7052-49c4-99e0-76a5e832977d-2012/?fullscreen&hints=off | https://xd.adobe.com/view/859c3a12-427f-4f08-9749-7d630c80ed7f-d0cf/?fullscreen&hints=off |
| SINGPASS | https://xd.adobe.com/view/ba2aa30c-b8da-40c4-aff7-35c39b92ed87-ef79/?fullscreen&hints=off | https://xd.adobe.com/view/b44a8363-e25f-4a98-8b35-d46d163a4709-e119/?fullscreen&hints=off |

## APPENDIX II

| **Code** | **Code Definition** |
|---|---|
| Avoidance | Users refrain from using information technology due to the fear of threats |
| Data sensitivity | Users Making choices based on the sensitivity of Data provided to the website |
| Source validity | Users check for the trustworthiness of the source via Google/Yahoo/etc Search when receiving the link |
| Personal device | Users trust the target website if accessed from a personal device or link saved in the device |
| Trust towards entity | Perceived trust towards the expected website provider |
| Checks security indicator | Looks at visual indicators such as URL, Lock Symbol or HTTPS |
| Checks aesthetics of website | Looks at the website's design, familiarity and possible grammatical errors to trust the website |
| Additional security | Adds 2FA security or other mechanism such as scanning tools |
| Check with Social circle | Users seek help from family members or friends |
| Check online | Users seek help from Search engines, online forums or social media |
| Mismatch in information provided | Mismatch in the designer's goals of conveying information and the user's understanding of the information presented. The information provided is not comprehensible to the user. |
| Lack of awareness | Doesn't understand/know about the existing protection mechanisms or online threats |
| Too much information | Presented with information more than necessary causing the user to feel overwhelmed |
| Incomplete information | Not able to make decisions based on the information provided |
| Misconception | Possessing an inaccurate understanding of security mechanisms/indicators work |
| Habituation | decreased response to an event due to repeated exposure |
| Trust reliability of helper | Trusts that the support giver would be able to identify a legitimate/phishing site |
| Coping strategy | Adopting an alternative strategy when not able to resolve a problem immediately or in a straightforward manner |
| Independent | Able to find out the validity of a site without external support due to technical ability |
| Contextual preference | Makes choices based on the condition such as the source of the link, the sensitivity of the transaction, website provider etc |
| Cannot trust reliability of helper | Doesn't't' trust that the giver of support would be able to help with identifying a legitimate/phishing site |
| Time taken to get response | Time is taken to get a response regarding the trustworthiness of a website matters |
| Availability of helper | State of being able to quickly get assistance from helpers for verifying website |
| Cause trouble for helper | The user feels that the act of asking for help might trouble the helper |
| Cannot trust reliability of helper | Doesn't trust that the giver of support would be able to help with identifying a legitimate/phishing site |
| Anxiety | User is afraid of being falling prey to cyber threats such as scams, data leakage, theft |
| Independent | Able to find out the validity of a site without external social support |
| Potential manipulation | Data might lack integrity due to potential manipulation by hackers or other malicious actors |
| Trusted towards platform | The outside social circle platform is run by a trusted entity or a trusted group of individuals |
| Reliability of Scoring mechanism | Users' feelings towards the scoring mechanism for representing the trustworthiness of a website. |
| Unfamiliar method | Unfamiliarity with technology or mechanism |
| Ai dependability | Trustworthiness and Reliability of AI in giving correct responses and the creator of the AI |
| Potential manipulation | Data might lack integrity due to potential manipulation by hackers or other malicious actors |
| Combination | Combination of multiple options |
| Wisdom of the crowd | larger groups of people can be collectively smarter in assessing the trustworthiness of the website |
| Availability and quick response | State of being able to quickly get assistance from AI, regardless of whether the response is accurate or not |
| Lack of preventive measures | Users who don't know how to keep themselves secure or are ignorant |
| Takes preventive measures | Users know how to keep themselves secure or demonstrate healthy cybersecurity habits |
| Vague - not appropriate for this question | NA/Not related to any of the research questions. |